\documentclass[sigconf]{acmart}
\usepackage{graphicx}
\usepackage{tikz, pgfplots, tikzscale}
\usetikzlibrary{fit, calc}
\usetikzlibrary{decorations}

\AtBeginDocument{%
  \providecommand\BibTeX{{%
    \normalfont B\kern-0.5em{\scshape i\kern-0.25em b}\kern-0.8em\TeX}}}

\copyrightyear{2020} 
\acmYear{2020} 
\setcopyright{acmcopyright}
\acmConference[ICSE-SEIP '20]{Software Engineering in Practice}{May 23--29, 2020}{Seoul, Republic of Korea}
\acmBooktitle{Software Engineering in Practice (ICSE-SEIP '20), May 23--29, 2020, Seoul, Republic of Korea}
\acmPrice{15.00}
\acmDOI{10.1145/3377813.3381348}
\acmISBN{978-1-4503-7123-0/20/05}

\begin{document}

\title{Building and Maintaining a Third-Party Library Supply Chain for Productive and Secure SGX Enclave Development}

\author{Pei Wang}
\authornote{Contributed equally to the paper.}
\email{wangpei10@baidu.com}
\affiliation{%
  \institution{Baidu Security}
  \city{Sunnyvale}
  \state{CA}
  \country{USA}
}

\author{Yu Ding}
\authornotemark[1]
\email{dingyu02@baidu.com}
\affiliation{%
  \institution{Baidu Security}
  \city{Sunnyvale}
  \state{CA}
  \country{USA}
}
\author{Mingshen Sun}
\email{sunmingshen@baidu.com}
\affiliation{%
  \institution{Baidu Security}
  \city{Sunnyvale}
  \state{CA}
  \country{USA}
}
\author{Huibo Wang}
\email{wanghuibo01@baidu.com}
\affiliation{%
  \institution{Baidu Security}
  \city{Sunnyvale}
  \state{CA}
  \country{USA}
}
\author{Tongxin Li}
\email{litongxin@baidu.com}
\affiliation{%
  \institution{Baidu Security}
  \city{Sunnyvale}
  \state{CA}
  \country{USA}
}
\author{Rundong Zhou}
\email{zhourundong@baidu.com}
\affiliation{%
  \institution{Baidu Security}
  \city{Sunnyvale}
  \state{CA}
  \country{USA}
}
\author{Zhaofeng Chen}
\email{chenzhaofeng@baidu.com}
\affiliation{%
  \institution{Baidu Security}
  \city{Sunnyvale}
  \state{CA}
  \country{USA}
}

\author{Yiming Jing}
\email{jingyiming@baidu.com}
\affiliation{%
  \institution{Baidu Security}
  \city{Sunnyvale}
  \state{CA}
  \country{USA}
}

\newcommand{\nlib}{159}

\renewcommand{\shortauthors}{Wang et al.}

\begin{abstract}
The big data industry is facing new challenges as concerns about privacy leakage soar.
One of the remedies to privacy breach incidents is to encapsulate computations over sensitive data within hardware-assisted
Trusted Execution Environments (TEE). Such TEE-powered software is called secure enclaves.
Secure enclaves hold various advantages against competing for privacy-preserving computation solutions. However,
enclaves are much more challenging to build compared with ordinary software.
The reason is that the development of TEE software must
follow a restrictive programming model to make effective use of strong memory encryption and segregation enforced by hardware. These constraints transitively apply to all third-party dependencies of the software. If these dependencies do not officially support TEE hardware, TEE developers have to spend additional engineering effort in porting them.
High development and maintenance cost is one of the major obstacles against adopting TEE-based privacy protection solutions in production.

In this paper, we present our experience and achievements with regard to constructing and continuously maintaining a third-party library supply chain for TEE developers. In particular, we port a large collection of Rust third-party libraries into Intel SGX, one of the most mature trusted computing platforms. Our supply chain accepts upstream patches in a timely manner with SGX-specific security auditing. We have been able to maintain the SGX ports of \nlib{} open-source Rust libraries with reasonable operational costs. Our work can effectively reduce the engineering cost of developing SGX enclaves for privacy-preserving data processing and exchange.
\end{abstract}



\maketitle

\section{Introduction}
Big-data technologies have enabled application vendors to rapidly push out innovative features and significantly improve service quality. The societal and economical benefits of analyzing large volumes of real-world data have led to personal data being collected and transferred at an unprecedented scale. Unfortunately, recently reported incidents show that the current practices of storing and managing personal data can be exceedingly insecure. With the introduction of new legislation like the EU General Data Protection Regulation (GDPR), user privacy is not only a moral responsibility but legal liability.

One of the effective ways to prevent privacy leakage without hurting data availability is to encapsulate computation over sensitive data within Trusted Execution Environments (TEE) where memory confidentiality is securely enforced by hardware. The enforcement does not rely on the cooperation of privileged software like operating systems and virtual machine managers, since they cannot be trusted due to conflicts of interest between users, service providers, and hardware platform owners.

To date, many chip makers have augmented their products with advanced TEE extensions. The Intel Software Guard Extensions (SGX) is one of the most mature TEE implementations on the market.
An application running in the SGX mode on an SGX-enabled CPU is called a \emph{secure enclave}.
An enclave is able to remotely prove its identity to clients, who will send their data to the application only if they are assured that
the data will be well protected by the enclave. Compared with other privacy-preserving solutions such as homomorphic encryption and federated learning, TEE is supported by specialized hardware, therefore much more efficient. At this point, it is generally much easier to support
realistic data processing workload with TEE-based software systems.

Unfortunately, developing SGX software turns out to be quite difficult due to various architectural constraints.
For example, SGX enclaves cannot directly perform I/O operations since the integrity and confidentiality of any resources
outside the SGX-encrypted memory are not guaranteed. In general, to enjoy the security and performance benefits of Intel SGX,
software developers have to adapt themselves to a much more restrictive programming model.
From the security point of view, it is imperative for software vendors to transit
to a more careful development paradigm if they aim to better protect data privacy. However, the SGX programming restrictions transitively apply to all
dependencies of an SGX program, meaning SGX developers cannot easily offload the development of commonly used functionalities by using third-party libraries. This drastically drives up the cost of developing complex SGX software, making software vendors reluctant to adopt the SGX technology.

In this paper, we share our experience with constructing and maintaining a supply chain of SGX-compatible third-party libraries
and making them readily accessible by SGX programmers, which helps alleviate the burden of developing SGX enclaves with complicated features.
Our supply chain offers open-source libraries written in Rust, an inherently memory-safe programming language with
its performance comparable to traditional system languages like C and C++. It has been widely acknowledged that Rust is one of the most suitable languages for SGX enclave development.

Starting from April 2018, we have been actively porting high-quality Rust libraries into SGX and consecutively synchronize them
with their upstream versions. We developed a comprehensive methodology to make sure that our ported libraries
comply with the SGX threat model. For libraries that are hard to meet this criterion, we inject API-breaking changes to force 
enclave developers to be well aware of the potential risks of relying on these libraries inside SGX.
To keep the maintenance cost at a manageable level, we built a pipeline to automate a large amount of the work needed to
update the forked libraries with upstream patches.
Our investigation shows that the third-party library supply chain has supported the development of numerous high-quality
open-source SGX projects. 

In summary, we make the following major contributions in this paper,
\begin{itemize}
    \item We identify the lack of third-party library support as a key software engineering challenge in building
    privacy-preserving big data applications with Intel SGX and provides a realistic solution to the problem.
    \item We propose a methodology to port third-party Rust libraries into SGX, making them comply with the secure computation
    thread model, setting up an accessible software supply chain to accelerate SGX enclave software development.
    \item We implement a highly automated pipeline to keep the cost of maintaining the SGX ports of \nlib{} third-party Rust libraries
    at a modest level.
    \item We demonstrate that our supply chain is already supporting real-world SGX enclave development. 
\end{itemize}

The remainder of this paper is organized as follows. Section~\ref{sec:background} introduces the background knowledge about
Intel SGX and Rust. Section~\ref{sec:objective} enumerates the objectives we would like to achieve in this work and the corresponding challenges.
Section~\ref{sec:structure} describes the overall structure of our library supply chain. We explain our methodology of building and maintaining the supply chain in Section~\ref{sec:methodology}. Section~\ref{sec:tooling} introduces our effort in providing additional tooling support
for SGX enclave development in Rust. We present empirical results in Section~\ref{sec:eval} to demonstrate the practicality of our methodology
and the real-world impact of our work. Issues and related work are discussed in Section~\ref{sec:discuss} and Section~\ref{sec:related}, respectively. We conclude the paper in Section~\ref{sec:conclusion}.

\section{Technical Background}
\label{sec:background}

\subsection{Intel SGX}
\label{sec:back-sgx}
SGX is Intel's latest major effort to enable trusted computing at scale. For each SGX-capable Intel CPU, a unique secret is
physically planted inside the chip during manufacture. Meanwhile, an asymmetrically paired secrete is retained by Intel to
verify the authenticity of the SGX CPU. At run time, the CPU reserves a special region of memory whose content is
constantly encrypted with keys derived from the planted secret. Furthermore, based on this secret, the chip can prove
to clients through remote attestation that it is running a program whose identity is known by the clients, after which the clients are
assured that their data will be well protected before uploading them to the program hosted by the SGX CPU. The trust 
between the chip and clients is not dependent on any intermediate software stack; instead, users only need to trust
Intel's ability to verify the authenticity of the SGX-capable CPU and the application software running in SGX mode.

From the perspective of SGX enclaves, system software is not trusted. This significantly reduces the size of required Trusted Computing Base (TCB)~\cite{Rushby:1981:DVS:800216.806586}; however, the resulting unprecedentedly restrictive thread model
also limits SGX programs' capability of acquiring computation resources from the outside untrusted environment.
In general, constraints faced by SGX developers include but not limited to the following,
\begin{itemize}
    \item Services provided by SGX enclaves can only be invoked through limited call gates pre-defined by enclave writers. These call gates are named \texttt{ECALL}s.
    \item Most hardware and software interrupts are not visible or available to SGX enclaves. That means SGX enclaves cannot directly
    request services from the OS, e.g., I/O and memory mapping. If OS-dependent resources are indispensable, the enclaves have to temporarily
    exit the SGX state through another set of pre-defined interfaces called \texttt{OCALL}s. 
    \item The \texttt{RDTSC} instruction is not supported in SGX. Trusted timestamp information can only be acquired by
    communicating with the Converged Security Engine which is part of the Intel Management Engine. This is several orders of magnitude slower
    than \texttt{RDTSC}, making it prohibitively expensive for certain applications.
\end{itemize}

Due to the constraints listed above, most existing software code is incompatible with SGX. To port legacy software into SGX, developers have to re-design the
trust boundary in their code, make sure the security sensitive part is adequately isolated, and define the \texttt{ECALL} and \texttt{OCALL} interfaces to bridge the trusted and untrusted parts of the software. If the partition is carelessly decided,
the enclaves can still be breached by the
outside malicious entities through Iago attacks~\cite{Checkoway:2013:IAW:2451116.2451145}. This makes developing and porting SGX programs a non-trivial software
engineering challenge.

\subsection{The Rust Programming Language}
Since the SGX hardware protects enclaves from direct interference of system software and other applications,
the actual security of an SGX enclave is mostly decided by the enclave's robustness of handling potentially malicious inputs.
According to a recent report, about 70\% of the vulnerabilities discovered in Microsoft products are due to memory safety issues~\cite{bluehat-il}. It is known that languages like C and C++, although widely used in system programming, can easily
introduce memory corruptions into the software even by experienced developers. 

Rust is considered to be a promising successor of C/C++ for developing security-sensitive software.
Like C and C++, Rust is statically typed and allows programmers to manipulate memory layout at a fine granularity with predictable performance.
More importantly, Rust by default is \emph{memory safe}.
Programming in Rust prevents most of the memory errors, including buffer overflow, dangling pointers, data races, and use of
uninitialized memory, etc. It can also detect some other programming errors like integer overflow, if specially configured.
Beyond memory safety and performance, Rust supports most expressive features found in other modern programming languages, such as
closures, generics, and pattern matching. Programming in Rust has been shown to be very productive~\cite{Anderson:2016:ESW:2889160.2889229}.

The implementation of Rust is open-source, with its compilers and other development tools actively maintained on GitHub. Rust developers have formed a vibrant open-source community, providing a rich collection of third-party libraries of high engineering quality. An open-source ecosystem
is crucial to SGX software development, since SGX enclaves are usually required to be audited by its users to establish trust.
Due to the aforementioned benefits, Rust is recognized as one of the most suitable programming languages for SGX software development.

Rust uses its own terms to express some of the common software engineering concepts. A Rust library is called a \emph{crate}. A tool called
\emph{\texttt{Cargo}} is used to manage crate dependencies for Rust projects. Rust also has a community registry for third-party crates called \emph{\texttt{crates.io}}, with which crate writers can publish their work for other developers to use.

\section{Objectives and Challenges}
\label{sec:objective}

\subsection{Enabling Rust Programming for SGX}
\label{sec:rust-sgx-sdk}
Due to the unique hardware features, SGX enclaves have an entirely different software dependency stack from that of the ordinary non-SGX programs.
Intel provides SGX-compatible standard libraries for C and C++, plus an SDK that allows programmers to utilize SGX functionalities
through C interfaces. However, there is no similar support for Rust. Hence, the first library we need to port and maintain
is the Rust standard library. There are some other Rust programming primitives not available in SGX, e.g., static variables, threading, and mutex.
We will also need to enable them.

\subsection{Porting Third-Party Libraries}
Porting a library into SGX can require extensive manual work, depending on its scale and complexity. Most third-party libraries have other
third-party dependencies. Therefore, the amount of work needed to port a single library can rack up to an unexpected level. Some libraries contain functionalities that are inherently incompatible with SGX. In those cases, we have to slice the libraries and keep only the
SGX-compatible parts. Human labor is usually mandatory in this process. Although there has been researched proposing methods to
automatically partition a software project to partially fit it into SGX~\cite{203211,wang2017binary}, none of them are mature enough for constructing and maintaining a high-quality software supply chain.

\subsection{Timely Updates}
Many libraries we ported are being actively developed. We aim to minimize the differences between their original versions and the SGX versions, especially regarding major functionality and security updates. This is a non-trivial task since the two versions of
the same library have forked. Even if the patches have been well tested for the original library, they may break in SGX. Also, some newly
added functionalities may not fully comply with the SGX threat model.

When the number of libraries included by our supply chain was small, we manually track updates from the upstream and manually merge
them into the SGX versions. This workflow quickly became impractical as the supply chain grows. Limiting the cost of manual inspection
has always been a challenge.

\subsection{Compatibility with Existing Development Practice}
The Rust community has its own tooling and automation settings for software development. On the SGX platform, we want to provide
Rust programmers with a development environment they are already familiar with. The majority of Rust developers heavily relies on the library manager \texttt{Cargo} (and its extension \texttt{Xargo}) and the community-run library registry \texttt{crates.io}. Our supply chain is committed to be compatible with these tools and facilities. For dual-purpose libraries, i.e., libraries that can be used either inside or outside SGX, we try to minimize the configuration effort needed for compiling them in different environments.

\subsection{Supply Chain Security}
The Heartbleed bug in OpenSSL is one of the most infamous examples of open-source community failure, showing how vulnerabilities in a widely subscribed software supply chain can be a major
security threat to the entire software industry~\cite{wang2015risk}. Recent analysis about several attacks on the JavaScript library manager \texttt{npm} has shown that malicious organizations and individuals can be actively undermining the security of open-source software supply chains~\cite{fsekeynote, npm, Konoth:2018:MIL:3243734.3243858}. For SGX, and privacy-preserving computation in general, the consequences of successful supply chain attacks may be even more disastrous. 
Preventing supply chain attacks is one of the decisive factors that drive us to maintain our own Rust library supply chain for SGX, partially isolated from \texttt{crates.io}.

\section{Structure}
\label{sec:structure}
As mentioned in Section~\ref{sec:rust-sgx-sdk}, there is no official support for developing SGX
software in Rust, even without considering the lack of third-party libraries. 
Therefore, as the first step towards productive SGX enclave development in Rust, we built a modified version of Rust standard library with SGX-specific functionalities like remote attestation integrated. This project, called Rust SGX SDK\footnote{\url{https://github.com/baidu/rust-sgx-sdk}}, serves as the foundation of
all third-party libraries we port and maintain in the supply chain. Rust SGX SDK is based on Intel's C/C++ SGX SDK. We have spent considerable efforts on the security, performance, and usability of Rust SGX SDK. In particular, we implemented a secure foreign function interface between Rust and C/C++. The design of the foreign function interface is formalized and proved to be memory safe. Details about this part of the work are discussed in a previous paper~\cite{rust-sgx-ccs}.

\begin{figure}[t]
    \centering
    \includegraphics[width=\linewidth]{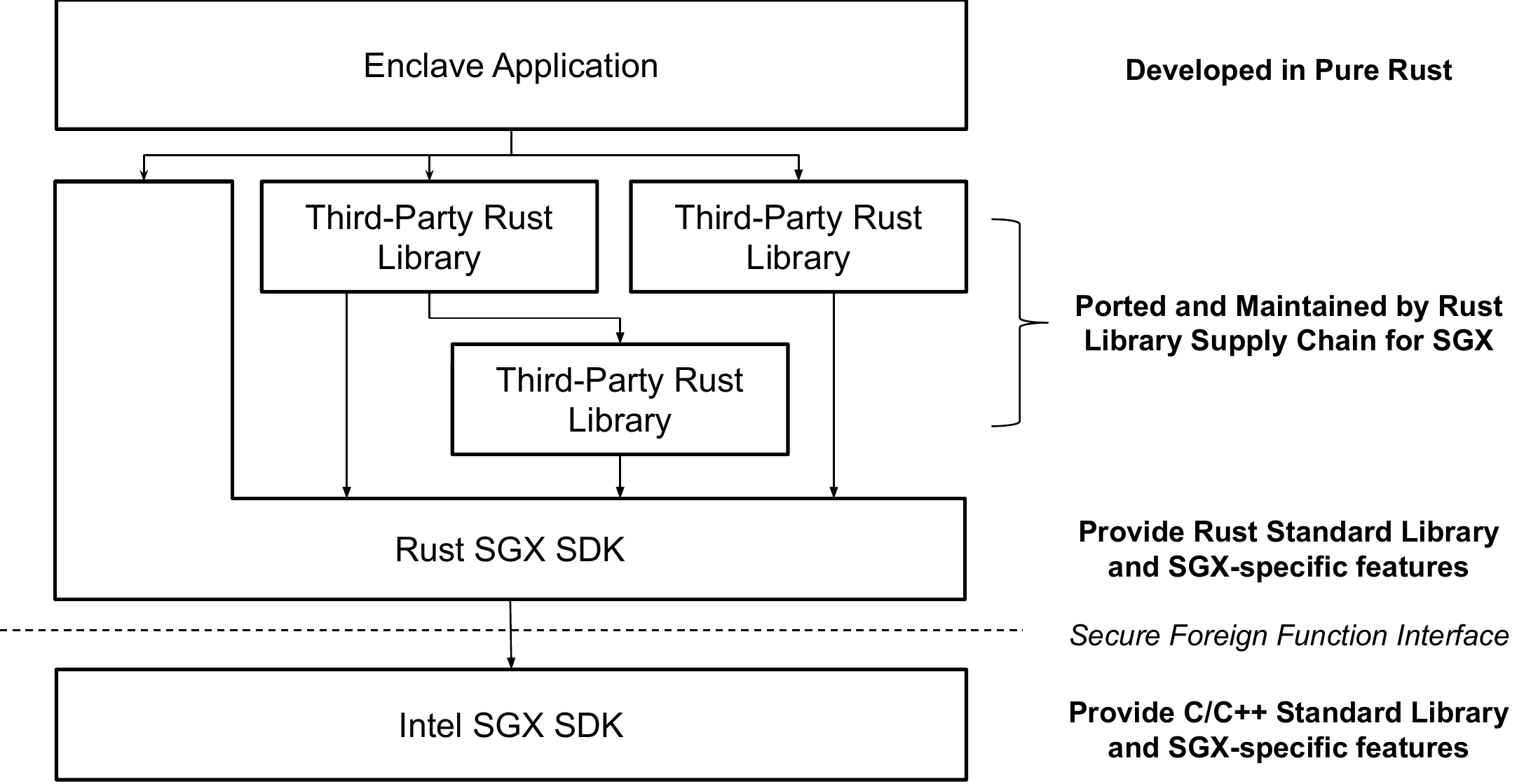}
    \caption{Structure of Third-Party Rust Library Supply Chain for SGX}
    \label{fig:overview}
\end{figure}

Figure~\ref{fig:overview} shows the structure of our supply chain. All third-party Rust libraries offered by the supply chain are dependent on the Rust standard library provided
by Rust SGX SDK, while Rust SGX SDK depends on Intel SGX SDK. Enclave developers cannot access raw C/C++ APIs provided by Intel. All enclave applications should be written in pure Rust to
minimize the chance of creating memory-based security vulnerabilities.

Some may view our dependency on Intel SGX SDK, which is written in C and C++, as a frailty that weakens our memory safety promise.
Indeed, it is possible that Intel's implementation becomes buggy and introduces vulnerabilities into the software stack, neutralizing
the benefits of using Rust for enclave development. However, implementing the functionalities provided by Intel SGX SDK in Rust leads to unreasonable engineering costs.

\section{Methodology}
\label{sec:methodology}
This section describes our methodology of constructing the Rust library supply chain for SGX. We report in detail how we decide
which libraries to maintain, how we port them into SGX, and how we keep the libraries synchronized with their upstream versions.

\subsection{Library Selection}
The first step towards building a software supply chain for SGX is to decide the roster of libraries to offer. This is a critical decision
because we are committed to providing long-term support for every library covered by our supply chain. We consider the following criteria when making a decision,
\begin{itemize}
    \item The library provides a functionality that is widely demanded in SGX enclave development, such as secure network protocols,
    cryptography, and data serialization.
    \item The library is of high code quality.
    \item The library promises or empirically keeps API stability.
    \item The library is a dependency of other admitted libraries and cannot be substituted by alternatives.
    
\end{itemize}
In general, a library has to meet several of the listed criteria to be included by our supply chain. That being said, we do not set up strict rules to regulate the admission but rather perform the selection on a case-by-case basis.
We constantly expand the supply chain by gradually porting more libraries to SGX. We identify new candidates based on our own development requirements and requests from the developer community. 

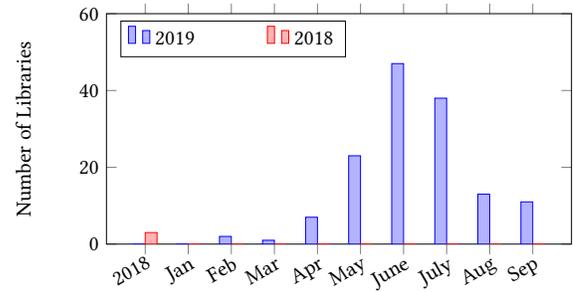
\begin{figure}[t]
\centering
\scalebox{1.0}{
\begin{minipage}{1.0\linewidth}
  \centering
  \resizebox{0.9\linewidth}{!}{
   \begin{tikzpicture}
 
\begin{axis} [
    ymin=0,
    ymax = 60,
    ybar,
    height = 5.0 cm,
    ybar=5pt, 
    try min ticks=5,
    ylabel={Number of Libraries},
    width = 1.0\textwidth,
    legend style={
                	{font=\normalsize},
                    at={(0.0,-0.15)},
                    anchor=north west,
                    legend columns=-1,
                    /tikz/every even column/.append style={column sep=1.0cm}
                 },
    symbolic x coords={2018,
                	Jan,
                    Feb,
                    Mar,
                    Apr,
                    May,
                    June,
                    July,
                    Aug,
                    Sep
                    },
    xticklabel style={rotate=30, anchor=north east, inner sep=0mm},
    legend pos= north west, 
    bar width=5pt,
    ybar=0pt,
    xtick=data
]
\addplot
  plot [error bars/.cd, y dir=both, y explicit]
  table [x=benchmark, y=native] {\mytable};

\addplot
  plot [error bars/.cd, y dir=both, y explicit]
  table [x=benchmark, y=paper] {\mytable};
\legend{2019, 2018}
\end{axis} 
\end{tikzpicture}
    }
    \caption{Number of Libraries Entering the Supply Chain in Each Month (Months in 2018 are consolidated)}
    \label{fig:fork-timeline}
  \end{minipage}%
  
  }
\end{figure}

We started constructing the supply chain in April 2018. By September 2019, the supply chain consists of
\nlib{} libraries. The supply chain has been expanding at a healthy pace. Figure~\ref{fig:fork-timeline} shows the timeline of its growth. We spent a considerable amount of time in porting the Rust standard library in 2018. Starting
from February 2019 when we finished migrating the standard library, our supply chain began to grow in a rapid speed.

\pgfplotstableread{
benchmark	native  paper
2018 -1  3 
Jan	0	-1
Feb	2	-1
Mar	1   -1  
Apr	7	-1
May	23	-1
June	47  -1		
July	38  -1
Aug	13  -1 
Sep 11  -1
}{\mytable}

\subsection{Porting Libraries into SGX}
The whole workflow of porting Rust libraries into SGX takes five major steps.
The first step is to inspect the dependencies of the to-be-ported library. The library manager of Rust can list the complete dependency graph for a library. Since Rust does not allow cyclic dependencies, we can always inspect one depended library at a time. We first analyze which of these dependencies should be stripped, e.g., dependencies used to support hardware or systems on which SGX is not available. For the remaining dependencies, if we can recursively port all of them into SGX, we can start porting the library itself. Otherwise, we have to abort the porting process. For the \nlib{} libraries currently included by our supply chain,
45 of them do not have extra dependencies that need to be ported and 132 of them have fewer than five such dependencies. The detailed breakdown is in Table~\ref{tab:dependency-dist}.
The statistics include transitive dependencies, but does not count dependencies that can be used for
 SGX enclaves without modification. Many of those dependencies are ``meta'' libraries that provide Rust syntax sugars, which
only take effects at the preprocessing stage of the compilation. They are mostly target independent and therefore do not need to be forked
by our library supply chain. 

\begin{table}[t]
\centering
    \caption{Ported Libraries Grouped by Size of Dependency Closure}
\begin{tabular}{cc}
\toprule
\textbf{Dependency Closure Size} & \textbf{\# of Libraries} \\
\midrule
0                                          & 45                                       \\ 
1                                          & 16                                       \\ 
2                                          & 19                                       \\ 
3                                          & 19                                       \\ 
4                                          & 13                                       \\ 
5                                          & 21                                       \\ 
6--10                                       & 12                                       \\ 
11--20                                      & 12                                       \\ 
$\geq 20$                          & 2                                        \\
\bottomrule
\end{tabular}
\label{tab:dependency-dist}
\end{table}


The second step is to pinpoint, in the library code, the usage of resources that have to be fetched from outside the SGX trust boundary, e.g., files, dates, and timezone information. We perform a thorough on the security implications of the usage of these resources. If using potentially untrusted resources does not lead to a security vulnerability, we implement a corresponding \texttt{OCall} to acquire the resources
from outside the SGX. Otherwise, we replace the untrusted resource with its trusted counterpart. For example, Intel provides a trusted file
system for SGX enclaves. The security of this file system does not rely on the OS but rather enforced by cryptographic algorithms with the keys
securely contained in SGX memory. Another example is that we can use the \texttt{RDRAND} instruction as the source of random numbers instead of OS-provided random number pools like \texttt{/dev/random}\footnote{There are security concerns about backdoors in hardware-provided random numbers. However, in the security model of SGX, Intel as the chip maker has to be trusted. Therefore, using \texttt{RDRAND} does not degrade
the overall security guarantees of SGX enclaves.}. Using this kind of trusted resources usually leads to un-neglectable performance cost, so we only make this
substitution unless absolutely necessary.

The third step is to disable multi-threading for the library.
Since SGX inherently distrusts the operating system, spawning new threads inside SGX enclaves is extremely challenging\footnote{In a sense, threads are special OS-dependent resources.}. There have been multiple solutions to this problem proposed by the industry~\cite{oe, asylo} and academia~\cite{234894}, but the general ideas are similar. When an SGX thread wants to spawn another, it first saves its context with an additional copy and exits the SGX state through a specially configured \texttt{OCall}; then the thread spawns a new thread outside SGX; after that, both threads enter SGX with a specially configured \texttt{ECall} and restore the previously saved context. Due to multiple CPU state switches in the process, the benefit of multi-threading is oftentimes overwhelmed by the cost. Therefore, we decide to completely disable multi-threading for libraries we maintain\footnote{It should be noted that our libraries are still reentrant. In other words, there can be multiple threads calling the same \texttt{ECall} interfaces to enter the SGX state simultaneously. We achieve this by supporting Rust
thread-local storage inside SGX. Rust itself does not yet support sub-function workload parallelism (e.g., splitting a loop to multiple threads), which is usually realized with the help of systems like OpenMP and Intel TBB.}. 

The fourth step is to port the test cases of the library. Unlike some other languages, Rust has a test framework bundled into the language.
This makes porting tests much less laborious since we only need to support one unit test driver for SGX. We first remove tests corresponding to functionalities that have been pruned from the library. We then collect
all tests into one enclave with a single \texttt{ECall} to speed up the testing process. Note that multi-threading inside tests are also
disabled.

The last step is feature simplification. Like C and C++, Rust supports conditional compilation.
Flags controlling the compilation conditions are called \emph{features}.
Library users can get slightly different variants of the same library by indicating what features are desired when compiling the library. One of the common scenarios where feature customization is used is that the library writer intends to explicitly optimize the code for different hardware architectures.
This kind of customization usually does not apply to users of our supply chain, since all users
are using very specific hardware, i.e., Intel CPUs supporting SGX. Therefore, we ``optimize out''
feature options not needed for SGX to make the usage and maintenance of the library less laborious.

\begin{figure*}[t]
    \centering
    \includegraphics[width=1\textwidth]{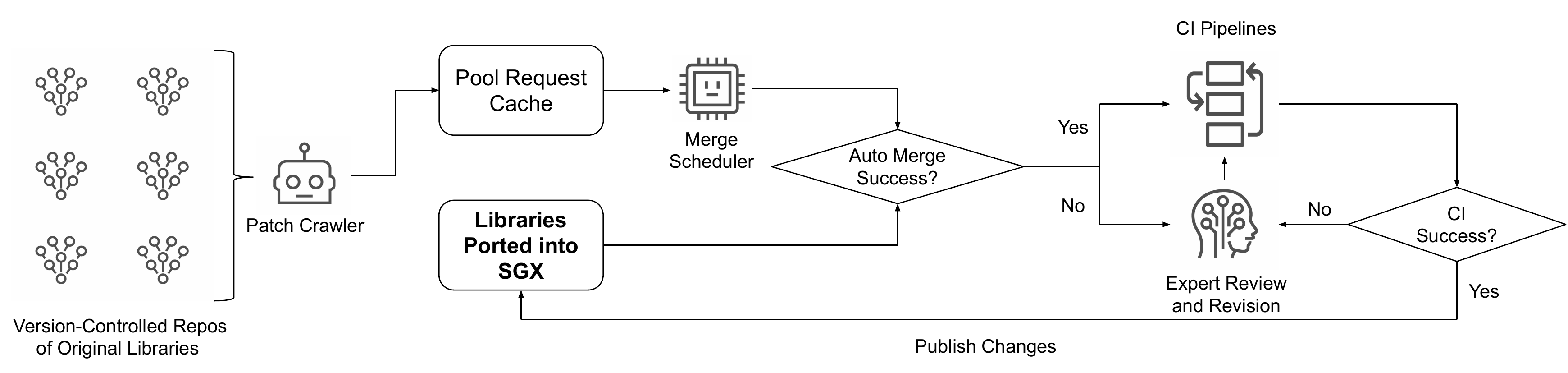}
    \caption{Overview of Supply Chain Update Infrastructure}
    \label{fig:automation}
\end{figure*}

\subsection{Patches and Updates}
\label{sec:patches}
We have built our own infrastructure to automate a large proportion of the work needed to keep our supply chain synchronized with
upstream patches. The infrastructure can be divided into three major components: the mega repository of all SGX-ported libraries, the pull request cache with a merge scheduler, and the continuous integration pipeline.

Figure~\ref{fig:automation} illustrates the workflow of our automation infrastructure. 
Currently, all of the libraries we ported are open-source and managed by Git. Our library supply chain takes the form of a mega Git repository
in which each SGX-ported library is a Git submodule, forked from its original repository. We devise a bot to monitor and collect patches submitted to the upstream repository. The collected patches will be sent to a pull request cache. The cache initiates automatic merge attempts based on decisions
made by a merge scheduler. For each library, the scheduler decides to start merging upstream patches with our SGX port when any one of the following conditions is met:
\begin{itemize}
    \item The commit message of a patch contains keywords like ``fix'', ``bug'', ``issue'', and ``release'', etc. 
    \item If the time elapsed since the last successful merge is over a threshold. Our current configuration for this duration is one month.
    \item If the number of patches accumulated in the cache has surpassed its capacity. By default, the cache capacity for a library is 10.
\end{itemize}

The automatic merge fails when upstream patches conflict with the code maintained by us. In this case, the failure is escalated to an expert in our team who will manually review and resolve the conflicts, until the patches can be successfully merged. After that, our continuous integration pipelines will run the test suite shipped with the updated library. If the CI test fails, a human expert will be notified to review the failure and perform necessary revisions to make sure the revised code passes the tests. The new versions are published once they go through all CI pipelines.

Note that for libraries we consider to be of extreme security significance, automatic merging is disabled. Every patch has to
be explicitly reviewed by an assigned library maintainer. Table~\ref{tab:critical-libs} lists and describes the functionality of those libraries. Indeed, the special treatment regarding those libraries imposes a considerable amount of human labor costs; however, we deem the cost as
necessary since software security is exceedingly critical for the SGX ecosystem. On the other hand, Rust is memory safe. From Our experience,
manually
auditing upstream updates for libraries listed in Table~\ref{tab:critical-libs} is completely manageable.

\begin{table}[t]
    \centering
    \caption{Libraries Requiring Mandatory Manual Review}
    \begin{tabular}{cc}
        \toprule
         \textbf{Library} & \textbf{Functionality} \\
         \midrule
         \texttt{rustls} & Transport Layer Security (TLS) protocol \\
         \texttt{webpki} & X.509 certificate validation \\
         \texttt{ring} & Cryptographic algorithms \\
         \texttt{cryptocorrosion} & Cryptographic algorithms \\
         \texttt{wasmi} & WebAssembly interpreter \\
         \bottomrule
    \end{tabular}
    \label{tab:critical-libs}
\end{table}

As previously mentioned, some of our libraries have external dependencies not included by our supply chain, since
they can be directly used by SGX enclave code without modification.
However, there is no guarantee that the future versions of those libraries will remain compatible with the ported libraries.
Therefore, in addition to the merge-driven CI tests, we routinely run CI tests every day. The
daily CI tests aim to capture those incompatibilities. When such an external library introduces breaking changes,
we either fork the library and include it into our supply chain or find a substitution for it.

The automation infrastructure significantly reduces the amount of manual work needed to maintain the supply chain. After this infrastructure
is online, a single maintainer with SGX expertise is able to manage all \nlib{} libraries we have ported. Indeed, the automation framework still leaves a portion of the work for human experts. However, we believe that part of the work should not be spared if we want to effectively
eliminate the risks of getting the supply chain compromised.

\subsection{Versioning}
Most library repositories offer different versions of the same library. Our SGX library supply chain does not follow this common practice. For most libraries, we only offer
the latest version.

The major reason behind this decision is that maintaining multiple versions of the same library
is not compatible with the Security Version Number (SVN) system of SGX. This special version number is used to control SGX-specific data backward compatibility. The SGX hardware ensures that data produced and sealed by enclaves with a newer SVN cannot be read by enclaves with an older SVN, even if they are signed by the same developer.

As displayed in Figure~\ref{fig:overview}, the entire supply chain is based on
Intel SGX SDK. When the SDK is updated to a new SVN, all SVNs of the libraries in the supply
chain should be updated accordingly. This could introduce non-linear expanding of SVN if we maintain different versions of the same library. For example, suppose we have a library \texttt{Lib} with
an initial version v1 and SVN $k$, built upon Intel SGX SDK with SVN $s$. At some point, \texttt{Lib} itself is updated to a new version v2 and SVN $k+1$, while the SVN of Intel SGX SDK remains $s$. Then at a later point,
Intel SGX SDK is updated to a new SVN $s+1$. Since Intel SGX SDK is the fundamental dependency,
its security properties propagate to all libraries in the supply chain and their SVNs
should be updated accordingly. However, we cannot handle this branching with a linear SVN. Suppose we update the SVN of \texttt{Lib}-v1 from $k$ to $k+1$ and update the SVN of \texttt{Lib}-v2 from $k+1$ to $k+2$, that would indicate \texttt{Lib}-v1 based on Intel SGX SDK of SVN $s + 1$ is compatible with \texttt{Lib}-v2 based on Intel SGX SDK of SVN $s$ in terms of security. But that is a false implication.

We have not developed a practical solution to this problem.
Potentially, we can maintain different copies of the supply chain, each based on a different version of Intel SGX SDK. But that would bloat the maintenance cost and lead to community fragmentation, making it much more difficult to assess and manage security risks.
Therefore, for now, we only keep the latest version of the same library in the supply chain.

\section{Tooling Support}
\label{sec:tooling}
Many Rust developers have their idiomatic tooling for writing and building the code. We found that some of the commonly used tooling is not available for SGX enclave developement, and the lack of them can notably affect engineering efficiency. Therefore, in addition to third-party libraries, our supply chain provides customized versions of development tools
that can be used for building SGX enclaves. This section enumerates our efforts in this aspect.

\paragraph{Protobuf}
Protocol buffers (a.k.a. protobuf) is a data exchange format developed by Google. It has become one of the most
popular data serialization solutions for open-source projects. To use protobuf, developers define the format of their messages using a
language- and platform-independent domain-specific language. A protobuf compiler is then invoked to generate the code stubs that can represent, parse,
and dump protobuf messages in the programming languages favored by the developers.
The Rust community has its own protobuf compiler implementation called \texttt{rust-protobuf}. Many libraries in our supply chain employ protobuf, but the code produced by \texttt{rust-protobuf} is not compatible with SGX. Therefore, we forked \texttt{rust-protobuf} and augmented it with
the capability of generating Rust code stubs that can be linked to SGX enclaves.

\paragraph{Code coverage}
Code coverage analysis is a widely needed software engineering capability for assessing the quality of test cases. Rust currently supports the LLVM code coverage analysis, and we spent effort to make it available for SGX enclaves as well. Our coverage analysis can merge results
from multiple threads and processes. By using our supply chain, enclave developers
can profile the code coverage of their integration tests for both the SGX part and non-SGX part.

\paragraph{Cross-Platform Optimization}
Since SGX is not available on all Intel CPUs, enclave developers often need to perform cross compilation.
Some SGX enclaves are very CPU intensive and developers want to utilize all possible optimization opportunities offered by the compiler.
This is hard to achieve using \texttt{Cargo}, the official library manager and compilation driver of Rust, since it offers very limited customization options for the target CPU, and the standard library is offered as pre-compiled binaries targeting the host CPU. 
To circumvent the constraints, some Rust developers use \texttt{Xargo}, a \texttt{Cargo} alternative that supports easy configuration of cross-compilation and allows them to rebuild the entire standard library. For example, the open-source deep learning compiler stack TVM~\cite{222575} relies on \texttt{Xargo} to build SGX enclaves.
Our supply chain has special support for \texttt{Xargo} cross-compilation targeting SGX-enabled CPUs. 

\section{Empirical Results}
\label{sec:eval}
We have gathered rich empirical data to demonstrate the practicality of our methodology and the value of our work. With these results, we aim to answer the following research questions,
\begin{itemize}
    \item[\bf RQ1] Does our library supply chain fulfill the need of SGX enclave developers?
    \item[\bf RQ2] What is the operational cost of building and maintaining such a supply chain?
    \item[\bf RQ3] Can our supply chain support real-world SGX enclave development?
\end{itemize}

\noindent {\bf RQ1: Does our library supply chain fulfill the need of developers?}

To attract a meaningful number of developers to subscribe to our library supply chain, we need to offer a wide range of selections to meet
various kinds of development demands. 


\begin{table}[t]
    \centering
    \caption{Supplied Libraries by Functionality}
    \begin{tabular}{cc}
         \toprule
         \textbf{Functionality Category} & \textbf{\# of Libraries} \\
         \midrule
        String Manipulation& 7\\
        Data Structure and Algorithm & 6\\
        Parsing & 3\\
        Binary Data Processing & 10\\
        Time and Date& 2\\
        Compression& 5\\
        Logging & 2\\
        Serialization& 11\\
        Randomness& 3\\
        Non-Cryptographic Hash& 20\\
        Image Processing& 5\\
        Crypto& 42\\
        Network& 7\\
        Safe Integer Processing& 8\\
        I/O & 2\\
        Scientific Computation & 2\\
        WebAssembly & 3\\
        Machine Learning & 1\\
        Blockchain Utils& 8\\
        Threading& 2\\
        Database & 2 \\
        Miscellaneous & 8\\
        \bottomrule
    \end{tabular}
    \label{tab:diversity}
\end{table}

We measure the diversity of the libraries by assigning each of them a functionality category. Table~\ref{tab:diversity} lists
the number of libraries in each category. It can be seen that our supply chain covers a wide range of commonly needed
functionalities in software development. In particular, we provide a rich collection of cryptographic libraries, which are
exceedingly important in privacy-preserving computation.  

\begin{figure}[t]
\centering
\normalsize
\resizebox{3.3in}{!}{%
	\begin{tikzpicture}
		\begin{axis}[
			ybar stacked,
			xlabel= {\huge Cut-off of most downloaded libraries on \texttt{crates.io} in 90 days (until Oct. 3, 2019)},
			ylabel = {\huge Percentage of each category},
			ymajorgrids = true,
			width = 0.9\textwidth,
			height = 0.5\textwidth,
			xmin = 0,
			xmax = 120,
			ymin = 0,
			ymax = 120,
			axis x line* = bottom,
			axis y line* = left,
			xticklabels = none,
			extra x ticks = {20, 40, 60, 80, 100},
			extra x tick labels = {Top 20, Top 40, Top 60, Top 80, Top 100},
			legend style = {at={(0.5, 1.025)}, anchor = south, legend columns = -1, draw=none, area legend},  
      area legend,
			scaled ticks = false,
			y tick label style = {/pgf/number format/use comma}
			]
			\addplot+[mark=none, fill, draw, bar width = 8mm]
			table[x=Rank, y=Ported] {\categorydata};
			\addplot+[mark=none, fill, draw, bar width = 8mm] table[x=Rank, y=Vendor] {\categorydata};
			\addplot+[mark=none, fill, draw, bar width = 8mm] table[x=Rank, y=Carrier] {\categorydata};
			\addplot+[mark=none, fill, draw, bar width = 8mm] table[x=Rank, y=Third-party] {\categorydata};
			\legend{Ported, No Need to Port, Not Applicable to SGX, Not Ported}
		\end{axis} 
	\end{tikzpicture}
	}
	\caption{Proportion of Most Downloaded Rust Libraries Covered by Our Supply Chain}
    \label{fig:popular}
\end{figure}
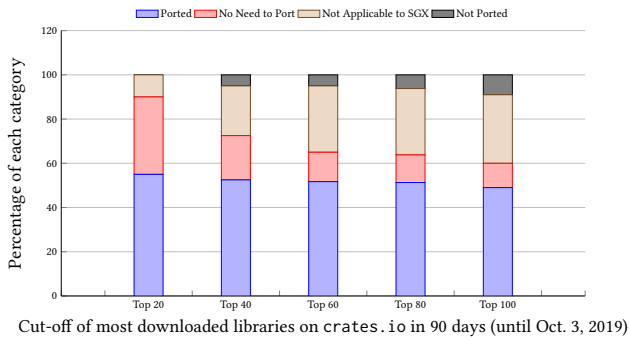

To show that our libraries can cover the need of a reasonably large population of Rust developers, we match the list of
our ported libraries with that of most downloaded Rust libraries hosted on \texttt{crates.io} in recent 90 days (until October 2019).
We divide libraries on \texttt{crates.io} into four categories,
\begin{itemize}
    \item Libraries we already ported into SGX.
    \item Libraries that can be directly used in SGX without modification.
    \item Libraries not applicable to SGX enclaves.
    \item Libraries that are potentially useful in SGX development but not ported yet.
\end{itemize}
The matching results are demonstrated in Figure~\ref{fig:popular}.
It can be seen that for the top 100 most popular libraries,
our supply chain can cover 60 of them (including libraries that do not need to be ported). Excluding libraries that are not applicable to SGX, only 9 libraries are not ported yet. If we consider the top 20 list, the availability rate is 90\% while the remaining 10\% is not applicable to SGX.

\pgfplotstableread[row sep=\\,col sep=&]{
    Rank	&	Ported	&	Vendor	&  Carrier	&	Third-party		\\
    20	&	55	&	35	&	10	&	0		\\
    40	&	52.5	&	20	&	22.5	&	5		\\
    60	&	51.7	&	13.3	&	30	&	5		\\
    80	&	51.3	&	12.5	&	30	&	6.2		\\
    100	&	49	&	11	&	31	&	9		\\
}\categorydata

%
%

We would like to emphasize that some of the libraries we ported make the development of certain types of enclaves significantly more productive.
For example, the TLS functionality provided by the \texttt{rustls} library allows networked enclaves to securely communicate
with clients, even if the TCP channels are provided by untrusted operating systems~\cite{knauth2018integrating}.
For another example, the \texttt{wasmi} library enables executing WebAssembly code in SGX, which is one of the
most wanted features by blockchain-related enclave developers.
In summary, we believe that our supply chain, with its current scale, can cover the need for a good portion of Rust developers.

\vspace{.1in}
\noindent \textbf{RQ2: What is the operational cost for maintaining such a supply chain?}

We configure each CI test to consist of 8 pipelines to cover different configurations, including two package managers (\texttt{Cargo} and \texttt{Xargo}), two OS versions (Ubuntu 16.04 and 18.04), and two build types (release and debug).
We assign a dedicated group of SGX-capable machines to our automation infrastructure for performing auto merge and continuous integration tests. By the time of writing, we have seven Intel NUC mini PCs with i7-8809G and five Lenovo SR250 servers with Xeon E-2186G. 

\tikzset{align at top/.style={baseline=(current bounding box.north)}}
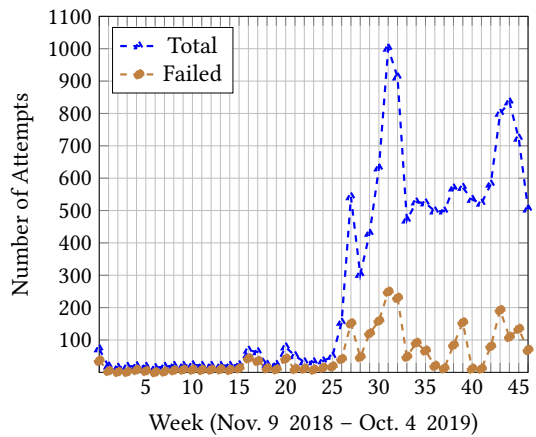
\begin{figure}[t]
  \centering
  \resizebox{.85\linewidth}{!}{
  \begin{tikzpicture}
    \begin{axis}[ 
    ylabel= Number of Attempts, 
    xlabel= Week (Nov. 9\, 2018 -- Oct. 4\, 2019), 
    xtick= {1,2,3,4,5,6,7,8,9,10,11,12,13,14,15,16,17,18,19,20,21,22,23,24,25,26,27,28,29,30,31,32,33,34,35,36,37,38,39,40,41,42,43,44,45,46,47,48,49,50},
    xticklabels = {,,,,,5,,,,,10,,,,,15,,,,,20,,,,,25,,,,,30,,,,,35,,,,,40,,,,,45,,,,,50},
    ytick = {100,200,300,400,500,600,700,800,900,1000,1100}, 
    yticklabels = {100,200,300,400,500,600,700,800,900,1000,1100}, 
    xticklabel style = {font=\Large}, 
    yticklabel style = {font=\Large},
    xlabel style = {font=\Large},
    ylabel style = {font=\Large},
    ymin = 0,
    ymax = 1100,
    xmin = 1,
    xmax = 47,
    legend style = {font=\Large}, 
    legend pos=north west, 
    grid=both]

      \addplot [line width=1pt, dashed, color = blue, mark=triangle] plot coordinates {
        (1, 72)
        (2, 19)
        (3, 15)
        (4, 18)
        (5, 20)
        (6, 18)
        (7, 14)
        (8, 18)
        (9, 21)
        (10, 21)
        (11, 23)
        (12, 21)
        (13, 21)
        (14, 21)
        (15, 21)
        (16, 23)
        (17, 68)
        (18, 63)
        (19, 24)
        (20, 24)
        (21, 78)
        (22, 51)
        (23, 32)
        (24, 32)
        (25, 35)
        (26, 49)
        (27, 154)
        (28, 542)
        (29, 302)
        (30, 429)
        (31, 631)
        (32, 1001)
        (33, 912)
        (34, 470)
        (35, 526)
        (36, 523)
        (37, 498)
        (38, 498)
        (39, 569)
        (40, 571)
        (41, 534)
        (42, 522)
        (43, 581)
        (44, 800)
        (45, 835)
        (46, 723)
        (47, 504)
    };
    
    \label{plot:Native}
     \addplot [line width=1pt, dashed, color = brown, mark=*] plot coordinates {
        (1, 36)
        (2, 4)
        (3, 1)
        (4, 1)
        (5, 7)
        (6, 4)
        (7, 0)
        (8, 2)
        (9, 6)
        (10, 8)
        (11, 8)
        (12, 7)
        (13, 8)
        (14, 9)
        (15, 7)
        (16, 12)
        (17, 43)
        (18, 35)
        (19, 12)
        (20, 9)
        (21, 42)
        (22, 10)
        (23, 11)
        (24, 8)
        (25, 14)
        (26, 18)
        (27, 41)
        (28, 151)
        (29, 47)
        (30, 120)
        (31, 159)
        (32, 249)
        (33, 230)
        (34, 48)
        (35, 91)
        (36, 67)
        (37, 20)
        (38, 12)
        (39, 84)
        (40, 154)
        (41, 11)
        (42, 11)
        (43, 80)
        (44, 192)
        (45, 109)
        (46, 134)
        (47, 69)
    };
    \label{plot:LibOSSGX}

     \addlegendentry{Total}
    \addlegendentry{Failed}

    \end{axis}
    \end{tikzpicture}
    }
  \caption{Auto Merge and CI Test History}
   \label{fig:ci-history}
\end{figure}


Figure~\ref{fig:ci-history} shows the weekly activities of our automation
infrastructure. Roughly, 15\% to 25\% of the auto-merge and CI test attempts failed. Most failures can be resolved in three days.
It should be noted that not all failures need to be handled manually.
For example, about 10\% of the CI test failures are due to network issues when there are too many CI tasks trying to connect to \texttt{crates.io} at the same time.
Occasionally, our daily CI tests fail for a large proportion of the libraries
in the supply chain due to an external dependency introducing
a breaking change. For example, we experienced 58 failures on July 16, 2019 because the \texttt{libc} library introduced a change that is incompatible with our supply chain, failing all CI pipelines that use
\texttt{Xargo} as the package manager.

\vspace{.1in}
\noindent \textbf{RQ3: Can our library supply chain support real-world SGX enclave development?}

We are aware that multiple commercial Rust-powered SGX products are depending on our supply chain for development. Unfortunately, we
are not able to disclose the details about these cases. Therefore, we choose to approach this research question by
investigating how the open-source community utilizes our supply chain to fulfill their development
requirements. All the repositories of the third-party libraries forked by our supply chain are hosted on GitHub. We do not get notified
when developers download these repositories, so it is difficult for us to get a comprehensive list of projects that make use of our supply
chain.

We take a best-effort approach to identify these projects. If a Rust project intends to depend on a library hosted by us, the developers
must provide the URL of the corresponding repository in the manifest file of the project. Based on this insight, we perform a code search on Github, using our organization account name as the keyword, since it is a common part of the URLs of all libraries in our supply chain. To prevent unprofessional projects from introducing bias into our analysis,
we apply a screening pass to the raw search results. In particular, we filter out projects without clear descriptions, documentations, or active commit histories. We also filter out projects for educational and training purposes.

\begin{table*}[t]
    \centering
    \caption{Open-Source Projects on Github Depending on Libraries from Our Supply Chain}
    \begin{tabular}{ccccc}
    \toprule
         \textbf{Project Name} & \textbf{Project Owner} & \textbf{Business Purpose}& \textbf{LoC (Rust/Total)} & \textbf{Dependencies} \\
    \midrule
         crypto-com/chain & Enterprise & Blockchain & 29,375 / 38,676 & 15\\
         enigmampc/enigma-core& Enterprise & Blockchain & 18,459 / 60,499 & 9\\
         provable-things/ethereum-keys-sgx& Enterprise & Blockchain & 1,802 / 2,322 & 7 \\
         smartcontractkit/chainlink-ios& Enterprise & Blockchain & 669 / 37,130 & 11 \\
         smartcontractkit/chainlink& Enterprise & Blockchain & 4,090 / 90,388 & 17 \\
         scs/substraTEE-worker& Enterprise & Blockchain & 6,049 / 11,808 & 29 \\
         scs/substraTEE-node& Enterprise & Blockchain & 9,493 / 10,148 & 4 \\
         mesalock-linux/mesatee& Community & Function as a Service & 21,298 / 58,867 & 39 \\
         dmlc/tvm& Community & Deep Learning Compiler & 5,686 / 414,247 & 2\\
         occlum/occlum& Enterprise & Library OS & 10,469 / 17,274 & 5 \\
         stenverbois/vulcan-rs& Individual & Automotive Control Networks & 944 / 7,336 & 4\\
    \bottomrule
    \end{tabular}
    \label{tab:client-projects}
\end{table*}

In the end, we identified a total of 11 qualified Github projects, listed in Table~\ref{tab:client-projects}.
We present their names, owner types, major business purposes,
the lines of code they contain, and the number of libraries from our supply chain on which they depend.

The majority of our users
work on projects related to blockchains and smart contracts, which is not surprising since they typically demand strong privacy guarantees for
their data. We also have users working in other areas, including deep learning and networked systems.
The scales of the projects are beyond trivial (up to 414K lines of code), even if we only consider the portion written in Rust (up to 29K lines of code).
We interpret the results as
strong evidence that our supply chain has the potential to support a diverse SGX software ecosystem.
We also would like to emphasize that most of the projects powered by our supply chain are enterprise products and reputed open-source projects,
showing that our effort has delivered meaningful impact.

\section{Discussions}
\label{sec:discuss}
\subsection{Lack of Automated Code Analysis}
For most software supply chains, security and engineering quality are the primary
concerns of their maintainers~\cite{EllisonSoftwareSupply2010,monteith2013exploring}. In our methodology, we mainly rely on
manual code review for quality assurance. However, historical research and experience emphasized the importance of automatic code inspection~\cite{EllisonEvaluatingand2010, boyens2015supply}, which are currently absent in our workflow.

At this point, we do not view the lack of automated code analysis to be problematic. On the one hand, the Rust type system
is already able to eliminate most low-level programming errors. To detect errors that evade the checks of the Rust compiler, such as
vulnerabilities caused by the infamous ``\texttt{unsafe}'' annotated Rust code, extremely sophisticated analysis techniques are required. There have been several prototypical techniques and tools
trying to tackle this problem~\cite{Jung:2017:RSF:3177123.3158154, baranowski2018verifying}, but none of them are ready for production.

On the other hand, security auditing for SGX code often requires a deep understanding of the unconventional threat model,
which is hard to formalize and difficult to be fed to automated program analysis systems.
Our experience is that even human experts can feel challenging to make the right decision about whether a piece of code should be pruned in the SGX version to prevent developers from misusing it.

We forecast that the current security auditing methodology can be improved as the trusted computing community gradually establishes the consensus about
appropriate engineering practices for TEE enclave development. Meanwhile, we have developed the prototype of
an automated tool to help code reviewers identify program locations that likely requires a security audit. This tool analyzes the call graph
of the enclave code and warns reviewers about the use of untrusted resources inside SGX.

\subsection{Scalability}
For the past 18 months, our supply chain has been built and maintained by a small team of three engineers.
For the long term, the concern is that as the supply chain keeps growing, will our current methodology still be sustainable?
A study on the scalability of Linux kernel maintainers' work~\cite{Zhou:2017:SLK:3106237.3106287} indicates that
a key factor affecting the workload of a maintainer is the number of files the maintainer works on. 
However, assigning additional maintainers for the same file only increases the productivity for a limited amount.
With more libraries to be included in our supply chain, we would have to increase the headcount of the maintenance team.
As a result, we expect that at one point in the future, we need to shift our maintenance model.

There are several potential directions to explore. The first is that we cap the size of the supply chain and only maintain
the most widely used libraries for SGX developers. If so, we can retain the current methodology.
The second possible direction is to merge the forked libraries back to the upstream so that the
support for the SGX target becomes official for these third-party libraries, although it
is not yet clear how security auditing should be conducted in this model. The third is to offload the maintenance work to
the open-source community and form a decentralized working group to direct the further development of the supply chain.
Nevertheless, for the latter two directions to be possible, we need to first expand the Rust SGX development community
and convince others that it is worthwhile to pay special attention to the problem. 

\section{Related Work}
\label{sec:related}
Besides our work, there are other efforts from the industry trying to build a productive SGX development community.
Google's Asylo~\cite{asylo} project provides a subset of POSIX APIs for developing SGX enclaves in C and C++. The Open Enclave project~\cite{oe} from Microsoft ported the standard C and C++ libraries into SGX, providing a more complete list of APIs than Intel SGX SDK. Both of the two projects try to abstract away the SGX-specific details such that they can be used to develop enclaves for other types of TEEs in the future. In general, Asylo and Open Enclave make it much easier to port third-party libraries into SGX since the transplantation no longer needs to redesign
the SGX-incompatible part of the code. The negative effects of this, in our humble opinion, is that it becomes too easy for SGX developers to access
potentially untrusted resources, which may loose the security consideration
in the development process. The Enclave Development Platform\footnote{\url{https://edp.fortanix.com/}} by Fortanix follows the same methodology and it supports Rust as we do. In general, our work has a different philosophy for SGX software development. We believe that it is more suitable for developers seeking the highest possible level of security for their enclaves.

The difficulty in porting software into SGX is also a well known academic challenge and a large volume of research has tried to improve the productivity and quality of the migration. Lind et al. developed a method to automatically dissect software source code into different parts, of which the security-sensitive ones are placed into SGX enclaves~\cite{203211}. Wang et al. proposed a similar technique for binary code~\cite{wang2017binary}. Liu proposed an automated way to transform real-time software into a TEE-compatible form while ensuring the software still meets the real-time demands~\cite{10.1145/3278122.3278137}. These methods require manual annotation on code that accepts or produces sensitive data and the security is enforced
at a rather fine granularity. For complicated data processing systems, this is usually insufficient.
Sinha et al. developed a compiler that can verify the confidentiality of memory addresses when the code is targeted for SGX~\cite{Sinha:2017:CVP:3106237.3106248}. SCONE~\cite{199364} and Graphene-SGX~\cite{203255} are containers that allow legacy
software to run as SGX enclaves without modification. They achieve this by simulating the programming primitives (mostly OS-dependent) that are missing inside SGX. Similar to work from the industry, these methods save the effort of porting individual programs, but they blur the trust boundary of SGX to the extent
that developers do not need to re-design their software to make it indeed privacy-preserving. Based on our porting experience, if unmodified, many applications will leak sensitive data into the untrusted environment or allow untrusted information to affect their behavior.

\section{Conclusion}
\label{sec:conclusion}
In this paper, we shared our experience with constructing and maintaining a third-party Rust library supply chain with auxiliary tooling to
assist SGX enclave development.
We described the challenges in developing SGX enclaves from a software engineering perspective.
We then explained our methodology of selecting suitable candidate libraries to port into SGX and how we can deliver
timely updates for the ported libraries with security assurance. In the evaluation, we showed that our supply chain offers an abundant enough collection of Rust libraries to ease the development of SGX enclaves for various business purposes. 
We believe that our work can encourage the big data community to employ SGX as a viable solution to privacy-preserving computation.

\bibliographystyle{ACM-Reference-Format}
\bibliography{ref}

\end{document}